\documentclass{aa}

\bibpunct{(}{)}{;}{a}{}{,}
\usepackage{graphicx,subfigure}
\usepackage{txfonts}
\usepackage{color, colortbl}

\definecolor{Gray}{gray}{0.9}

\begin{document}
   \title{An approach to the analysis of SDSS spectroscopic outliers based on Self-Organizing Maps}
   
   \titlerunning{An approach to the analysis of SDSS outliers based on SOMs}

   \subtitle{Designing the Outlier Analysis software package for the next Gaia survey}

   \author{D. Fustes \and M. Manteiga \and C. Dafonte \and B. Arcay
          \inst{1}
          \and A. Ulla\inst{2}
          \and K. Smith\inst{3}
          \and R. Borrachero\inst{4}
          \and R. Sordo\inst{5}
          }
          
   \authorrunning{D. Fustes et al.}

   \institute{Universidade da A Coru\~na (UDC)\\
              \email{dfustes@udc.es ,manteiga@udc.es, dafonte@udc.es, cibarcay@udc.es}
              \and
              Universidade de Vigo (Uvigo)\\
              \email{ulla@uvigo.es}
              \and
              Max Planck Institute For Astronomy (MPIA)\\
              \email{smith@mpia-hd.mpg.de}
              \and
              Universitat de Barcelona (UB)\\
              \email{rborrachero@am.ub.es}
              \and
              Osservatorio Astronomico di Padova (INAF)\\
              \email{rosanna.sordo@oapd.inaf.it}
             }

   \date{\today}

% \abstract{}{}{}{}{} 
% 5 {} token are mandatory

\abstract
  % context heading (optional)
  % {} leave it empty if necessary  
   {}
  % aims heading (mandatory)
   {A new method is applied to the segmentation, and further analysis of the outliers resulting from the classification of astronomical objects in large databases is discussed. 
The method is being used in the framework of the Gaia satellite DPAC (Data Processing and Analysis Consortium) activities to prepare automated software tools that will be used to derive 
basic astrophysical information that is to be included in Gaia final archive.}
  % methods heading (mandatory)
   {Our algorithm has been tested by means of simulated Gaia spectrophotometry, which is based on SDSS observations and theoretical spectral libraries covering a wide sample of astronomical objects. Self-Organizing Maps (SOM) networks are used to organize the 
information in clusters of objects, as homogeneous as possible, according to their
 spectral energy distributions (SED), and to project them onto a 2-D grid where the data structure can be visualized.}
  % results heading (mandatory)
   { We demonstrate the usefulness of the method
 by analyzing the spectra that were rejected by the SDSS spectroscopic classification pipeline and thus classified as ``UNKNOWN''. 
Firstly, our method can help to distinguish between astrophysical objects and instrumental artifacts. Additionally, the application of our algorithm to SDSS objects of unknown nature 
has allowed us to identify classes of objects of similar astrophysical nature. In addition, the method allows for the potential discovery of hundreds of novel objects, such 
as white dwarfs and quasars. Therefore, the proposed method is shown to be very promising for data exploration 
and knowledge discovery in very large astronomical databases, such as the upcoming Gaia mission.}
  % conclusions heading (optional), leave it empty if necessary 
   {}

%\keywords{Gaia mission, unsupervised classification, Self Organizing Maps, Outliers, astronomical archives}


   \maketitle
%
%____

\section{Introduction}
The ESA Gaia mission, which is now in phase D (Qualification and Production), is expected to be launched by September 2013. It will provide 
the first highly accurate 6-D map of the Milky Way, measuring positions, parallaxes, and motions to the microarcsec level. The satellite's complex
 instrumentation, mode of operation, astrophysical main objectives, and its expected scientific performance have been extensively reviewed elsewhere, see for example 
\cite{SciencePerformanceGaia}. Since Gaia is the first non-biased survey of the entire sky down to approximately magnitude 20, it is raising enormous expectation from a wide range of astronomical research areas, going from Solar System to Cosmology, especially after it was decided that the final archives, containing the observations and basic astrophysical 
products, will be made public immediately after being produced.
 
The spacecraft will measure every object in the sky over 80 epochs on average and over the course of its 5 years of operating time, allowing for variability studies as well as an increase in 
the signal-to-noise ratio with time. We expect approximately $10^{12}$ observations and an extensive number of iterations, which will be required to process astrometry, 
photometry, and radial velocities, with the additional challenge of a flux of data of approximately $10^{5}$ observations per second \citep{HollandLindegren12}.

The main astrophysical properties of astronomical objects observed by Gaia will be derived by a software pipeline, which
 is being produced by an international consortium, the Gaia Data Processing and Analysis Consortium (DPAC). DPAC arose, in response to an ESA Gaia Announcement of Opportunity in 
March 2007, as an international collaboration with memberships from all over Europe, which nowadays includes a community of over 400 scientists and software engineers from more than 20 countries. DPAC is organized in several coordination units (CUs) and
responsible for a well-defined set of tasks in the Gaia data processing effort. CU8 was in charge of classifying the observed astronomical sources by both supervised 
and unsupervised algorithms, and of producing an outline of their main astrophysical parameters \citep{LNEA2010}. CU8 is subdivided into several work packages: DSC 
(Discrete Source Classifier) is the main package for classification, whereas GSP-Phot (General Stellar Parameterizer - Photometry) and GSP-Spec (General Stellar Parameterizer - 
Spectroscopy) are the main parameterization packages.
There are a number of additional packages dedicated to more specific tasks, such as Quasar/Galaxy parameterization (QSOC and UGC, respectively) or specific stellar population 
parameterizers (ESP). Finally, there are two packages dedicated to the unsupervised analysis of the raw data, OCA (Object Cluster Analysis)
and OA (Outlier Analysis). OA is the package that is described in this work, aimed at analyzing classification outliers.
   
\begin{figure}
\centering
\subfigure[Normalized passbands of Gaia instruments as a function of wavelength. G stands for Gaia broad-band white light, RVS refers to Gaia's Radial Velocity Spectrograph, 
and BP and RP refer to blue and red spectrophotometers]{ \includegraphics[width=\hsize] {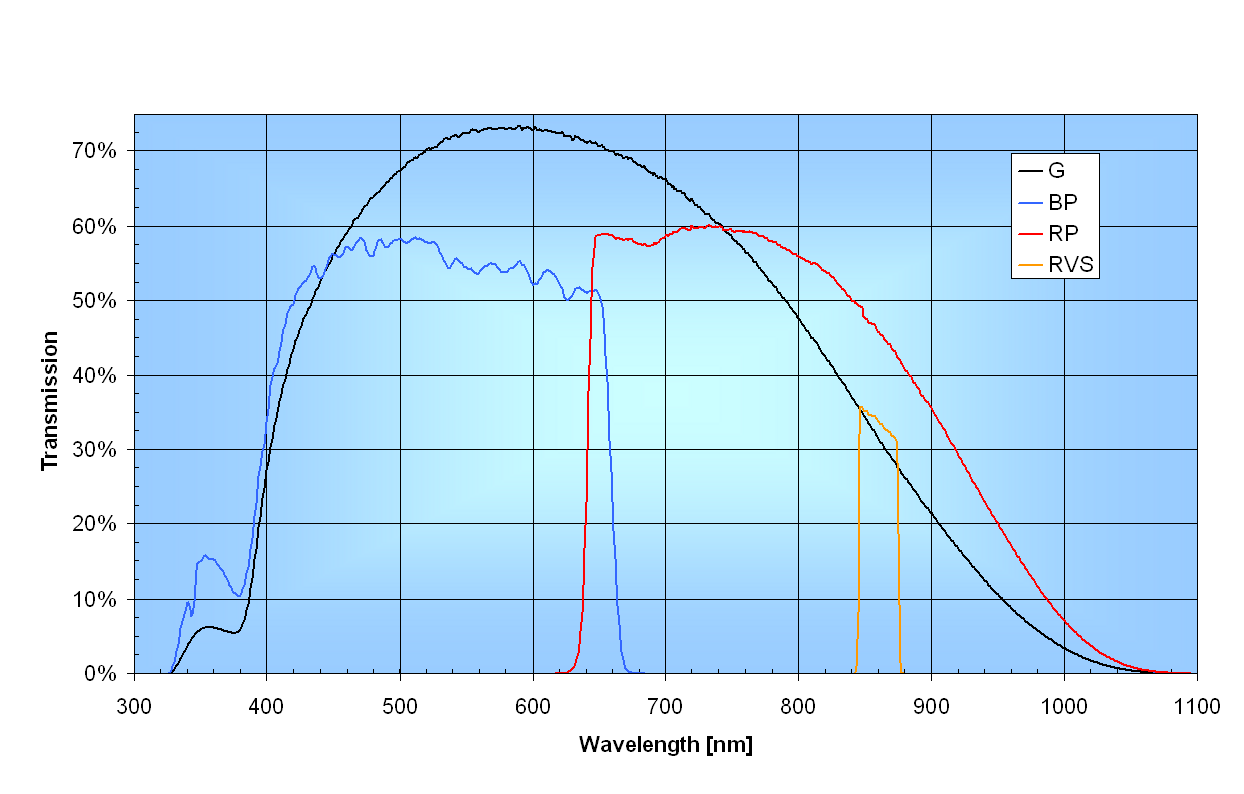} \label{fig:passbands} }
\subfigure[Spectral dispersion of Gaia BP and RP spectrophotometers]{ \includegraphics[width=\hsize] {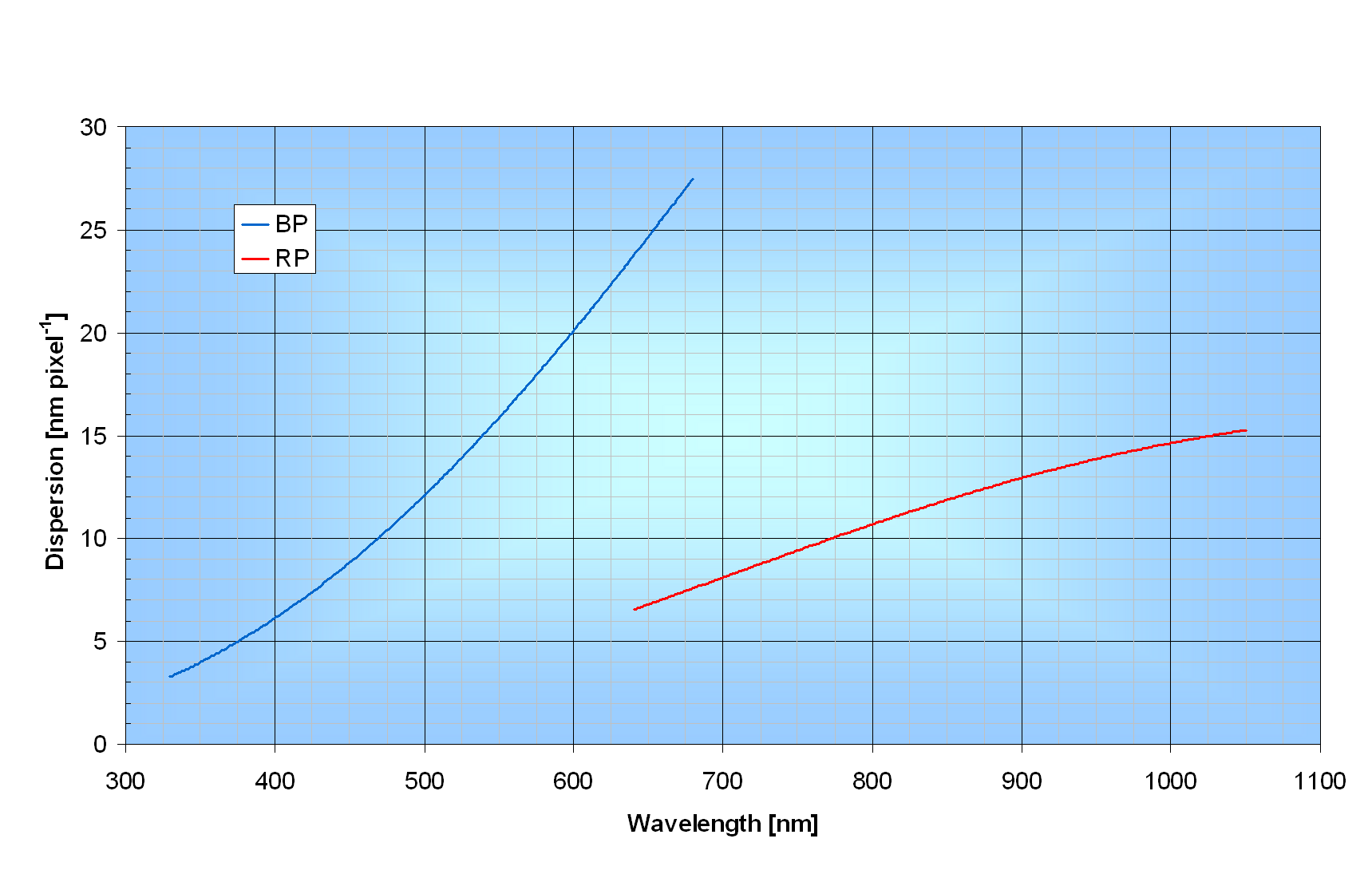} \label{fig:dispersion}}

\caption{Gaia spectrophotometers BP and RP properties. Credits: ESA.}
  \label{fig:BPRP}
\end{figure}

CU8 \textbf{Astrophysical parameters inference system (Apsis, \cite{AA/2013/22344}) is composed by a number of algorithms} to derive information on the nature of all astronomical objects that will be observed by the satellite,
mainly through the analysis of their astrometric properties and their Spectral Energy Distribution (SED). The SED will be obtained for all objects observed by Gaia, using two spectrophotometers: BP 
(Blue Photometer, operating in the wavelength range of 300-680 nm) and RP (Red Photometer, range 640 to 1050 nm). Figures \ref{fig:passbands} and \ref{fig:dispersion} show 
the normalized passbands as the instrumental response to photons, and the spectral dispersion as a function of wavelength and for BP and RP instruments.

The main objective of DSC consists in providing a probabilistic classification for every object observed by Gaia from among a well-defined set of astronomical classes: STAR, WD, PHYSICAL BINARY, GALAXY, QSO, and NON-PHYSICAL BINARY (composite object). White dwarf stars are considered an additional class (WD), so that
STAR includes all non-WD single stars. DSC is mainly based on a supervised method, more specifically a SVM (Support Vector Machine, \cite{SVM95}) algorithm working on Gaia spectrophotometry, which is complemented by other two subclassifiers working with astrometric data, see \cite{DSC2012} for further information.

OCA is using an unsupervised classification algorithm, called HMAC, see \citep{HMAC}, with the objective to determine the ``natural'' observed classes of astronomical objects among all Gaia observations, without any a-priori hypothesis about their physical nature. The expectation of CU8 is that both
 general classification working packages will be able to classify, with a reasonable level of reliability, approximately 95\% of all
  Gaia observations. 

Our group has been participating in the CU8 DPAC activities since 2007, and is responsible, among others subjects, for analyzing the classification outliers that result from DSC and OCA operations. The knowledge of the SED of each astronomical object, together with its precise astrometry and information about variability will, undoubtedly, provide us with the most complete physical information to describe its astrophysical nature. However, it is important to stress that, although the physics of the stars is nowadays well understood and there are extensive archives containing information about the light distribution of most known astronomical objects, it is expected that Gaia will observe such an enormous amount of sources that many of them could differ significantly from model predictions or previous observations. Gaia will observe a significant sample of peculiar objects, such as supernovae, stars with abnormal
 abundance patterns, Wolf-Rayet stars, multiple systems, or high redshift quasars, as well as, probably, new kinds of previously unseen objects. In addition, low signal-to-noise ratios, cosmic rays, instrument artifacts, and other damaged data will eventually occur, leading to classification errors. To deal with this issue, DSC is using an automatic outlier detector, based on a one-class SVM, that rejects objects that are far from the training data space (see for example, \cite{Scholkopf2001}). It is estimated that approximately $5*10^{7}$ objects (5\% of the total) will be marked as UNKNOWN by the DSC outlier detector, which means that some type of automatic analysis becomes mandatory. Furthermore, some objects will receive a set of probabilities that is not decisive in terms of final classification, so that their nature should be clarified by further analysis. These
 objects will be processed by OA.

The remainder of this paper is organized as follows: Section \ref{sect:simulations} presents the data that is being used to test the CU8 algorithms, Section \ref{sect:oaalgorithm} describes the algorithm used to process the outliers with OA
 and the resulting performance with Gaia simulations, and Section \ref{sect:sdssoutliers} presents the results that were obtained by the
 algorithm working with 10,125 spectroscopic outliers from SDSS (Sloan Digital Sky Survey), which may lead to the identification
 of new astronomical objects. Finally, Section \ref{sect:conclusions} outlines our conclusions and discusses the adaptation of the proposed
  methodology for data mining in the next Gaia mission. 
  
\section{Gaia simulated libraries}\label{sect:simulations}
DPAC is using a powerful simulator, the Gaia Object Generator (GOG,  \citep{gog}), to simulate a wide variety of observations that are
 foreseen to take place during the mission. Among the data generated for testing CU8 algorithms, we used a number of spectra from the SDSS Data Release 7 \citep{SDSSDR7}, transformed by 
GOG to BP/RP low resolution format and instrumental characteristics. We shall refer to such spectra as the SDSS Semi-empirical libraries, which are actually 3 libraries containing different
classes of objects: stars, quasars, and galaxies \citep{GalaxiesSE2012}. In addition, in order to complete the set of reference spectra, we considered model-based BP/RP DPAC libraries, 
 which were compiled in different ways from stellar synthetic spectra obtained from MARCS and PHOENIX models \citep{PhoenixGaia} and \citep{MarcsGaia}.  We also considered compiled libraries
 composed by binary stars \citep{WhiteDwarfs2006}, white dwarfs, ultra cool dwarfs \citep{UCD2000},
emission line stars, and planetary nebulae \citep{PN}. Finally, spectra for non-physical pairs were generated by adding the spectra of either two stars, a star and a galaxy, or a star 
and a quasar. The Gaia stellar libraries were presented and compared in \cite{Sordo2011}.

\begin{figure}
\centering
\subfigure[A quasar spectrum from SDSS]{\includegraphics[width=\hsize] {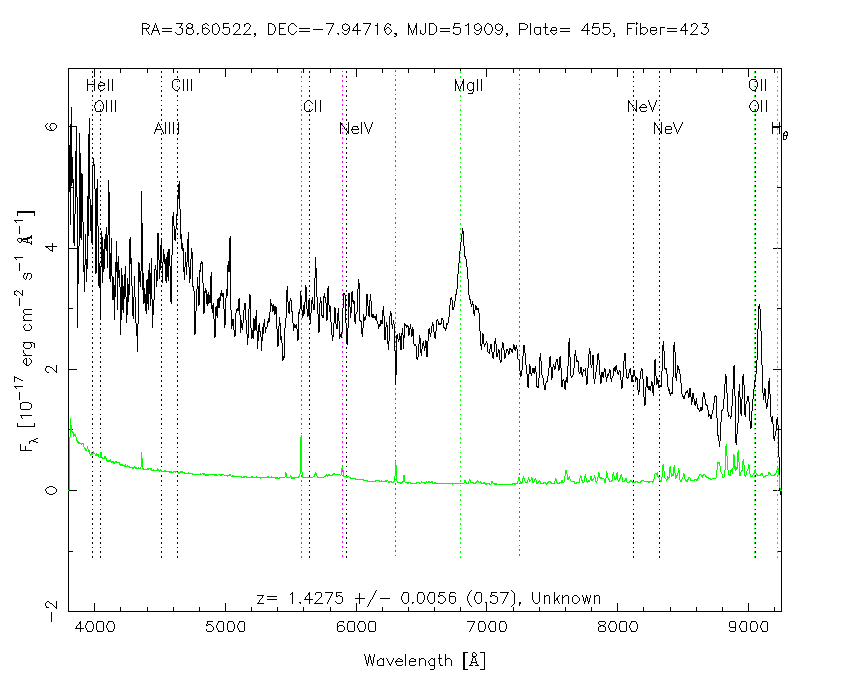} \label{fig:qsoSDSS} }
\subfigure[Quasar spectrum after simulating it with GOG]{\includegraphics[width=\hsize] {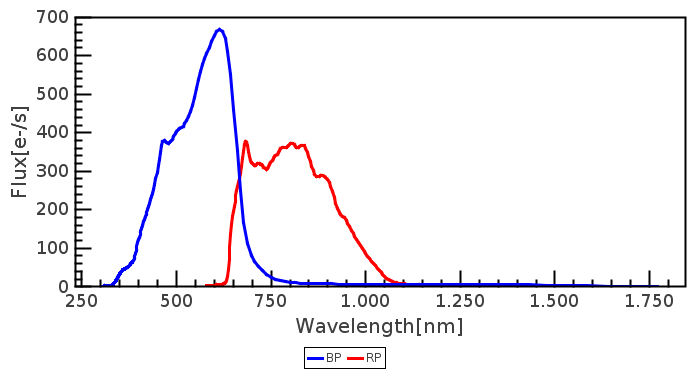} \label{fig:qsoGaia} }
\caption{Comparison between a SDSS spectrum for a QSO and its GOG simulated BP/RP spectrophotometry.}
  \label{fig:qso}
\end{figure}

Simulated Gaia spectrophotometry, delivered by DPAC to CU8 and used by Apsis, is currently produced without any correction for instrumental sensitivity: the true response function will not be 
known until late in the mission, since it will be the result of photometric calibration using standard stars. This is the reason why the available spectrophotometry is mostly 
dominated by a low frequency signal showing a typical belt structure for each of the photometers. Figure \ref{fig:qso} presents a SDSS spectrum for a QSO, and its GOG version 
(BP/RP spectrophotometry). GOG spectra are internally calibrated (i.e. the flux, magnitudes, and colors are correct). Section \ref{sect:sdssoutliers} discusses the impact of using 
this representation as opposed to working directly with SDSS spectra.  \textbf{During operations Gaia DPAC will need to run a regular assessment of the effect of bandwidth nonuniformities
by comparing accumulated spectra of sources of similar spectral type taken in different CCD rows and at different times, see \cite{2013Fabricius}.
It is expected that external calibration should be able to remove this.}

Both the astronomical classification and the main parameters of the objects populating the previous datasets are well known, and the compiled libraries extensively cover 
the range of physico-chemical and evolutionary parameters expected for the main classes of astronomical objects, thus allowing for the use of this data to test or 
validate our method for identifying ``classical'' object categories (see Section \ref{sect:oaalgorithm}). However, our algorithm is aimed at processing objects with an unknown 
nature, such as instrumental failures, faint observations, etc. Keeping this in mind, we compiled a new library formed by spectra from SDSS that were classified as 
``UNKNOWN'' by the SDSS spectroscopic classification pipeline. After removing some spectra with zero or negative fluxes, we obtained a dataset composed by 10,125 objects, mostly faint 
objects (with a mean magnitude of 19 in G band),
 incomplete spectra, and unsuccessful observations. The performance of our algorithm in the SDSS outliers library is discussed in Section \ref{sect:sdssoutliers}.

\section{Outlier Analysis algorithm}\label{sect:oaalgorithm}

Any study related to the classification of extensive datasets has to address the problem of analyzing multidimensional classification outliers.
 Among others, this problem has recently arisen in the analysis of astronomical surveys such as Pan-STARRS1, \citep{PanSTARRS-Classification}, the
  Sloan Digital Sky Survey, SDSS, \citep{SDSS-GalaxyClassification}, and the Blanco Cosmology Survey \citep{BlancoCosmology}. Since outliers are, 
by definition, objects that do not fit in the existing models, the analysis of large outlier datasets must be done by means of unsupervised
 algorithms, which do not consider any knowledge a priori. In the Data Mining field, there are two main approaches to deal with multidimensional
  data based on unsupervised techniques: Dimensionality Reduction and Clustering. Dimensionality Reduction tries to reduce the number of dimensions
  (variables, attributes) in the dataset to a level where they can be more reasonably analyzed by domain experts. Principal Component Analysis
  (PCA, see \cite{PCA}) is the best known algorithm of this kind. On the other hand, Clustering is aimed at grouping the data into a number of clusters that share similar properties. 
A wide variety of clustering algorithms has been proposed, as it is an ill-defined problem, see \cite{surveyClustering}; \cite{surveyFuzzyClustering}; \cite{surveyTimeSeries}.
  
Our choice for analyzing outliers is a clustering algorithm based on Self-Organizing Maps (SOM, \cite{SelfOrganizingMaps}). SOM have been used extensively in a number
 of scientific fields. Indeed, the paper that opened the field, \cite{Kohonen1982}, currently counts with more than 5000 citations. However, 
 they have been used sparingly thus far in Astronomy (\cite{Naim1997}; \cite{Geach2011}; \cite{Way2012}).

The main advantage of the SOM is that they provide quality clustering and non-linear dimensional reduction at the same time, by projecting
the data into a fixed number of clusters (called neurons or units in the Neural Networks field), arranged in a 2D (or 3D) structure, usually a matrix
with N rows by M columns. Each cluster has a representative, called prototype %(neuron weights in the Neural Networks field), 
which is a virtual pattern that better represents or resembles the set of input patterns belonging to such a cluster. The problem to be optimized is to find the best prototypes 
for the SOM clusters. Since this is a NP-hard problem, an iterative optimization procedure is followed to reach an acceptable solution from a randomly selected initialization 
of neuron weights. Firstly, for each input pattern, the neuron which most resembles the pattern is activated. This is calculated by means of the squared euclidean distance between the 
pattern and the neuron prototype. Then, the activated neuron and its neighbours are updated according to the activating patterns. The number of neurons in the neighbourhood of the 
activated neuron is large in the first iterations, but shrinks as the iterations succeed themselves. In this way, the algorithm starts sorting out the neurons and then smoothly moves to focusing on 
the clustering procedure, minimizing the residual (also called quantization error) between the neuron prototype and its activating patterns.

We have carried out several experiments with Gaia spectrophotometry and SOM (see \cite{hsc2012} and \cite{FustesSOM2012}), with considerable success. The experiments show that different object 
classes lie in well-defined map regions, and that it is possible to compress the data objects in a reduced number of clusters
 without significant loss of astrophysical information. See, for instance, the distribution of the objects on the map in Figure 3 
%and the confusion matrix in Table \ref{tab:cleanConfMatrix}, 
where we computed a 30 by 30 SOM (900 clusters) for 150,417 objects simulated with GOG, covering a wide variety of astronomical classes with varying parameters
described in Table \ref{tab:classes}. A confusion matrix is a useful tool to evaluate the success of a classification algorithm. It is a table in which each row represents 
an objects class, for which we compute the percentage of objects falling into clusters where the predominant class corresponds to each of the colums. The confusion matrix 
corresponding to the SOM clustering of the previous experiment is presented in Table \ref{tab:cleanConfMatrix}. The last row shows the number of objects per class in the input dataset.
In this case, the achieved compression rate is 167:1, with a mean class purity around 98.5\% in the SOM clusters. 

Note that it was necessary to apply some preprocessing to the BP/RP data before presenting it to the SOM. We started by joining both BP and RP in a single vector, which was then  
normalized to have a unit area, otherwise the SOM would only focus on apparent magnitudes. BP and RP present a wavelength region of overlapping and different spectral sensitivity (see
Figure 2).
We performed tests with two BP/RP formats, one using the two spectra that just merged one after the other, hence showing a central region with several 
near-zero points, and a second one with spectra obtained by matching the overlapped wavelength region between both photometers. In the first case we preserve the information on the 
spectral colors, but introduce several pixels that correspond to the same wavelengths. It has also been taken into account that low pixel values will not significantly affect 
the performance of the SOM neurons. Tests carried out with both data configurations allow us to confirm that only small differences were found, approximately 3\% in clustering purity, which 
proves that the small color differences are not introducing significant changes in the obtained groups. We decide to use merged spectra without wavelength redundancy, hereby saving some computation time.

\begin{table}
\caption{Classes of objects among Gaia simulations used for the testing of OA algorithms.}          
\label{tab:classes} 
\begin{tabular}{|c|c|}
\hline 
\textbf{CLASS} & \textbf{DESCRIPTION} \\ 
\hline 
AFGKM & Main sequence stars, from PHOENIX model \\ 
\hline 
OB & Very hot OB stars \\ 
\hline 
WD & White dwarfs, both DA and DB \\ 
\hline 
UCD & Ultra cool dwarfs stars \\ 
\hline 
GALAXY & Semi-empirical galaxies \\ 
\hline 
QSO & Semi-empirical quasars \\ 
\hline 
PN & Semi-empirical planetary nebulae \\ 
\hline 
\end{tabular}
\end{table}          

\begin{figure}
\centering
\includegraphics[width=\hsize]{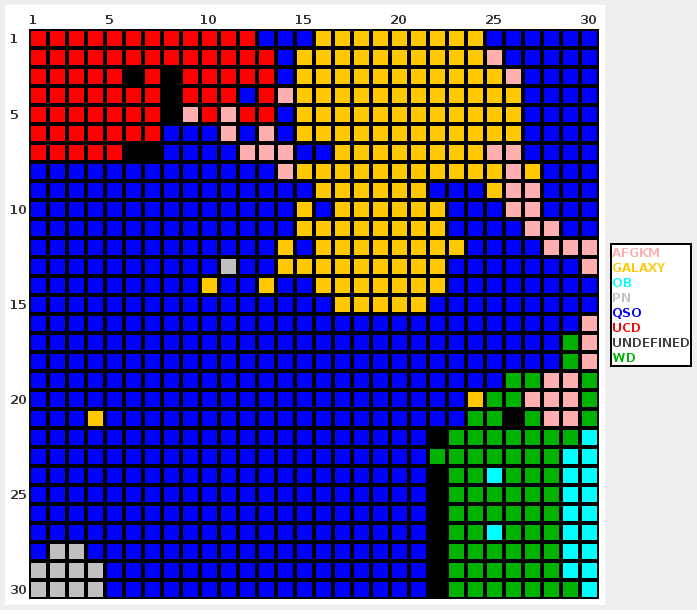}
\caption{Distribution of astronomical object classes obtained with Gaia photometric simulations, over a computed SOM, with 30 by 30 clusters. The color assigned to each of the cluster
 was set in function of the predominant class in the objects belonging to it. The black color indicates that the cluster is empty.}
\label{fig:cleanlabels}
\end{figure}

\begin{table*}
\caption{Confusion matrix of the SOM computed using Gaia simulations of a wide variety of astronomical objects, as explained in Section \ref{sect:oaalgorithm}. }             % title of Table
\label{tab:cleanConfMatrix}      % is used to refer this table in the text
\centering                          % used for centering table
{%
\newcommand{\mc}[3]{\multicolumn{#1}{#2}{#3}}
\begin{center}
\begin{tabular}{|r|r|r|r|r|r|r|r|}
\hline
 & \textbf{AFGKM} & \textbf{GALAXY} & \textbf{OB} & \textbf{PN} & \textbf{QSO} & \textbf{UCD} & \textbf{WD} \\ \hline
\textbf{AFGKM} & 95,28 & 1,3 & 0 & 0 & 0 & 2,7 & 0,72 \\ \hline
\textbf{GALAXY} & 0,67 & 98,4 & 0 & 0 & 0,94 & 0 & 0 \\ \hline
\textbf{OB} & 0 & 0 & 99,76 & 0 & 0 & 0 & 0,24 \\ \hline
\textbf{PN} & 0 & 0 & 0 & 99,33 & 0 & 0 & 0,67 \\ \hline
\textbf{QSO} & 0,44 & 0,83 & 0 & 0 & 98,7 & 0,10 & 0,10 \\ \hline
\textbf{UCD} & 0,51 & 0 & 0 & 0 & 0 & 99,49 & 0 \\ \hline
\textbf{WD} & 1,46 & 0 & 0 & 0 & 0 & 0 & 98,54 \\ \hline
\rowcolor{Gray}
 &  &  &  &  &  &  &  \\ \hline
\textbf{COUNT} & \multicolumn{1}{r|}{5000} & \multicolumn{1}{r|}{33670} & 9999 & 748 & 70554 & \multicolumn{1}{r|}{9890} & \multicolumn{1}{r|}{20556} \\ \hline
\end{tabular}
\end{center}
}%

\end{table*}                 
 
In general, when Gaia satellite observations are available, we do not know the physical nature of the objects that populate the obtained SOM clusters. Therefore,
an identification phase will be mandatory, in order to provide at least a description of the cluster by all available means, including both Gaia internal data such as known SEDs,
objects variability, astrometry, photometry etc., and external data, such as information from other astronomical surveys, 
possibly complemented by human experts knowledge and additional ground observations when necessary. The topology preservation of the SOM can help in this sense, since it 
provides researchers with meaningful visualizations, such as the well-known U-Matrix, that serve as maps for data exploration (see \cite{DataExplorationSOM}). 
Section \ref{sect:sdssoutliers} puts into practice these visualizations, and others specifically designed for the task, to unveil the nature of the objects populating the 
SDSS spectroscopic outliers.

\section{Unveiling the nature of SDSS outliers with the OA algorithm}\label{sect:sdssoutliers}
This section describes the processing of outlier analysis by means of the abovementioned SOM algorithm. To do so, we shall apply our method
to a set of data of unknown nature, concretely the SDSS outlier library described in Section 2. We shall try to identify the classes of astronomical sources that populate it. 
This case is realistic, in the sense that the physical nature
of the objects can be considered unknown since they were rejected by a classification pipeline. The following sections describe the process of
computation of a SOM that is well-suited for this dataset, and a posterior identification procedure of the clusters populating it. 

\subsection{SOM learning procedure}
The first step in the analysis of the SDSS outliers is to set the learning parameters for the SOM. Some of them can be fixed with simulations after some experimentation,
such as the number of learning iterations and the neighbourhood function. But the most important parameter to set, the size of the map
(the number of clusters) is more difficult to estimate, since it strongly depends on the data. We opt for using a measure of error in the clustering, called Mean Quantization Error 
(MQE), to stablish the map size \citep{SOMQuality}. MQE measures the mean distance among a cluster prototype and the objects populating it. As such, we established that 30 by 30 is
 an acceptable
map size for the present experiment. Finally, we selected the batch learning mode instead of the online mode, because the batch mode has the advantage of being independent from the order in which the 
patterns are presented to the SOM \citep{SOMBatch}.

\begin{figure}
\centering
\includegraphics[width=\hsize]{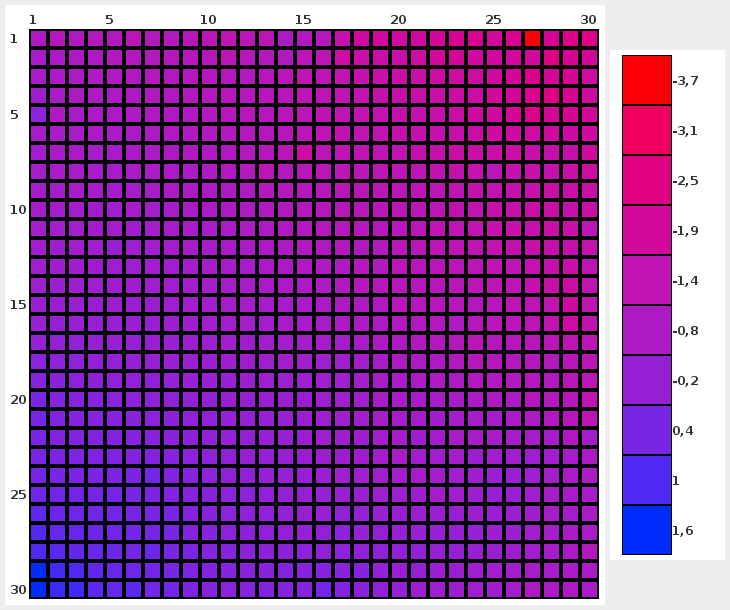}
\caption{Photometric color distribution $G_{rp}-G_{bp}$ (magnitude differences in integrated BP and RP bands) in the SDSS outliers SOM.}
\label{fig:SDSSColor}
\end{figure}

The MQE index measures the quality of the clustering, but we still do not know if the map is correctly orderedd according to the input topology. It is difficult to estimate this mathematically. One way to assess the ordering is to visualize the color ($G_{rp}-G_{bp}$) distribution (magnitude differences in integrated RP and BP bands) in the SOM prototypes,
as shown in Figure \ref{fig:SDSSColor}, where a general ordering in the color distribution can be visualized. Additionally, 
with this plot and the MQE we were able to select the best SOM among several randomly initialized learning procedures.   

\subsection{Data navigation through SOM visualizations}
Once the final SOM is obtained, several visualization tools are available with which to unveil the data's physical nature and distribution. For instance,
the color plot described in the previous section can be used as a guide for guessing stellar atmospheric temperatures. 

\begin{figure}
\centering
\subfigure[Distance<71.9E-4]{\includegraphics[width=0.45\hsize] {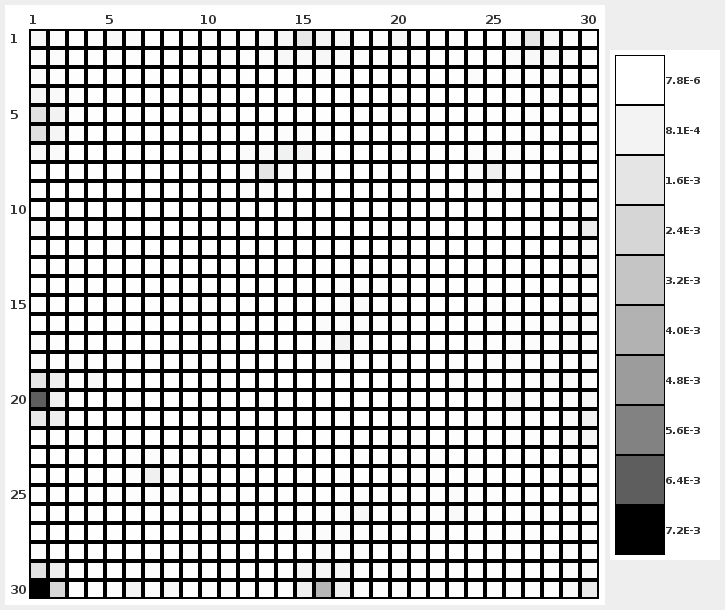} \label{fig:umatrix} }
\subfigure[Distance<13.8E-5]{\includegraphics[width=0.45\hsize] {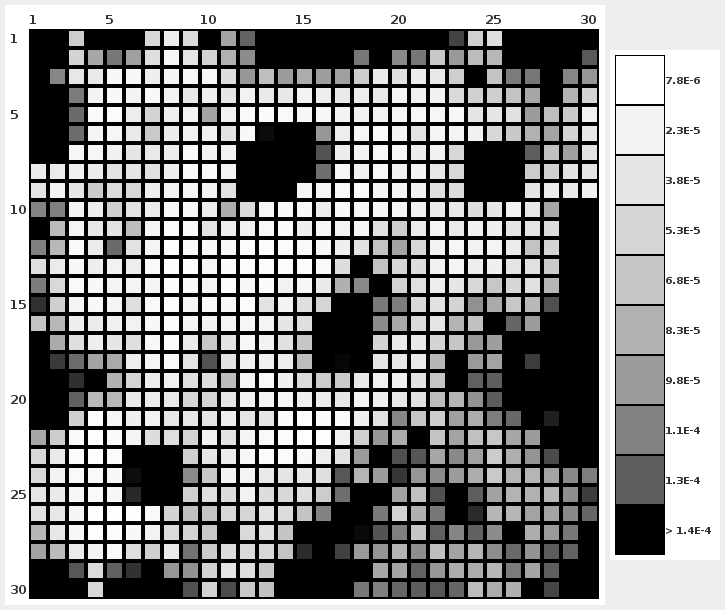} \label{fig:umatrix20} }
\subfigure[Distance<80.8E-6]{\includegraphics[width=0.45\hsize] {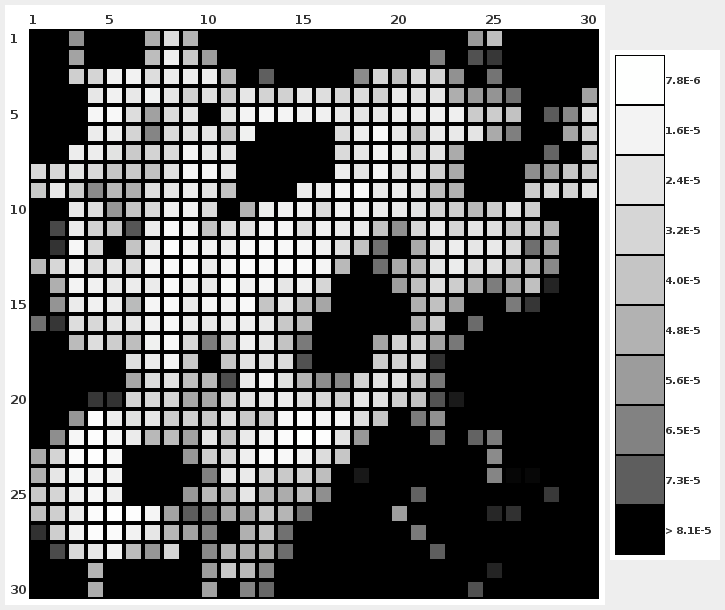} \label{fig:umatrix40} }
\subfigure[Distance<39.8E-6]{\includegraphics[width=0.45\hsize] {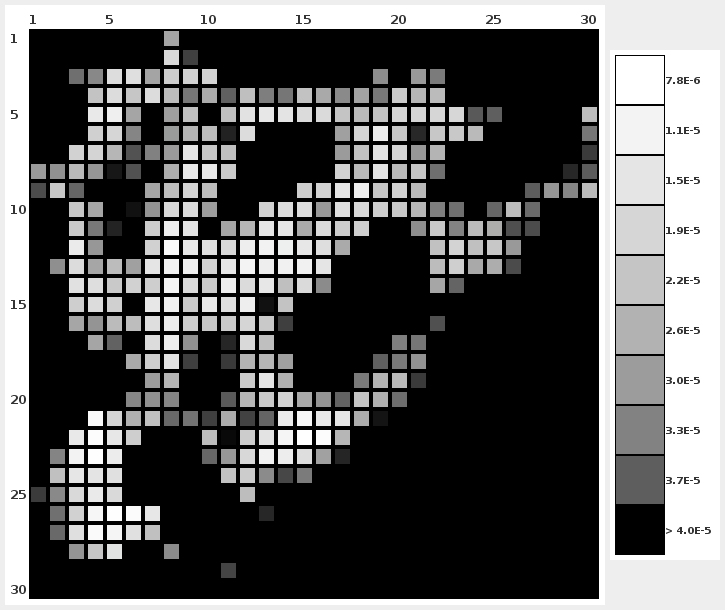} \label{fig:umatrix60} }
\caption{U-Matrix for the outlier SOM. The different plots correspond to different distance limits, which allow us to change the image contrast so as to unveil the underlying structure.}
  \label{fig:UMatrix}
\end{figure}

The prototypes of each cluster in the SOM can also be visualized, provided that there is an expert that is capable of interpreting the BP/RP data in some way. Such an exploration can be 
performed region by region in the SOM instead of visualizing every cluster, since clusters that are located close in the SOM are close in the input space as well. 
The U-Matrix, which displays the distances among clusters, is a visualization tool that can
assist the expert in the analysis process. The distance between the adjacent clusters is calculated and presented with different gray levels. 
A dark color corresponds to a large distance and thus a gap between the clusters in the input space, whereas a light color between the clusters means that they are close to each other 
in the input space. Light areas can be thought of as dense regions in the input space, whereas dark areas correspond to more sparse ones. Fig. \ref{fig:UMatrix}
shows the U-Matrix computed from the SOM built for the SDSS outliers dataset, at several levels of contrast. This way, we can identify clusters
that are outlying with respect to the others, such as the darker one in position (30,1). In addition, we can select several groups of clusters for joined identification. 
These groups can be further studied in order to propose a taxonomy among the unclassified objects in the SDSS survey, as will be shown in the next sections. 

The nature of the outliers (darker clusters) detected in the U-Matrix can vary: they can be instrumental errors, bad detections, or faint or unexpected objects. The analysis of 
the darker regions in the map has to be carried out in a more detailed way (even cluster by cluster), since the objects in each of the dark regions could differ considerably from each other.

\subsection{Identification of SOM clusters using spectral templates}\label{sect:labeling}
\begin{figure}
\centering
\subfigure[QE<92.0E-4]{\includegraphics[width=0.45\hsize] {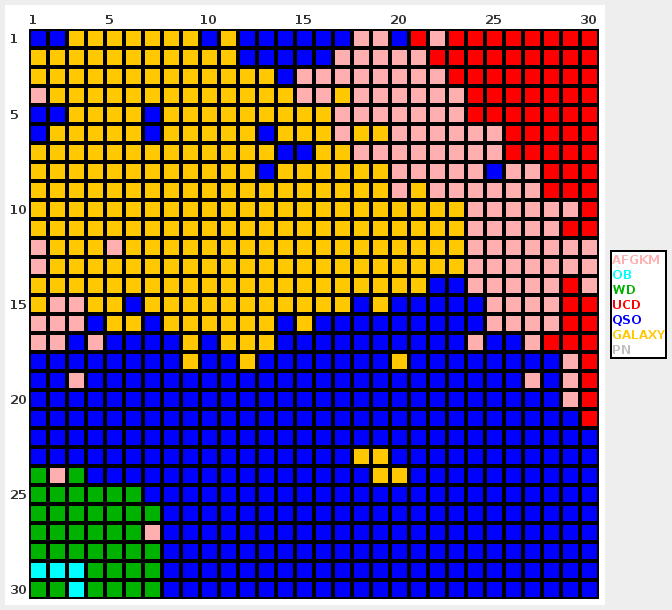} \label{fig:templateModes} }
\subfigure[QE<34.7E-5]{\includegraphics[width=0.45\hsize] {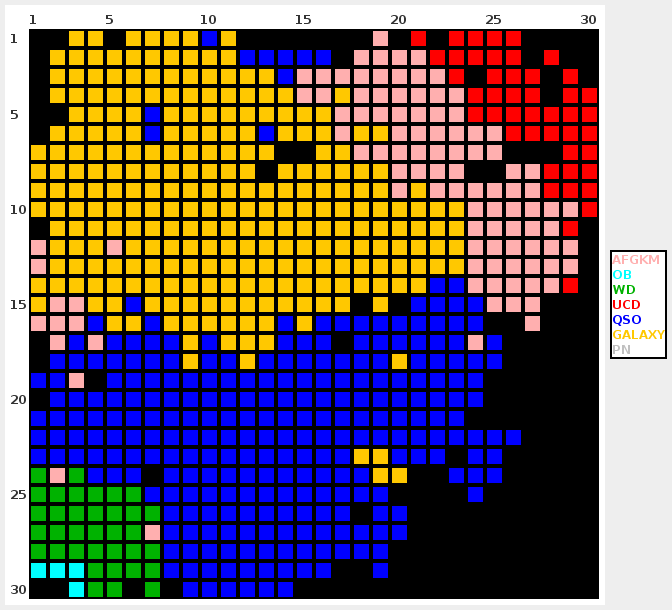} \label{fig:templateModes20} }
\subfigure[QE<14.3E-5]{\includegraphics[width=0.45\hsize] {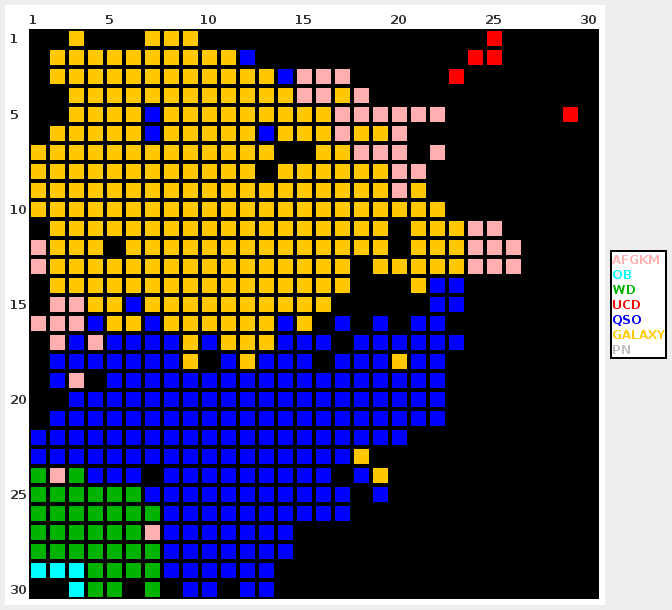} \label{fig:templateModes40} }
\subfigure[QE<53.3E-6]{\includegraphics[width=0.45\hsize] {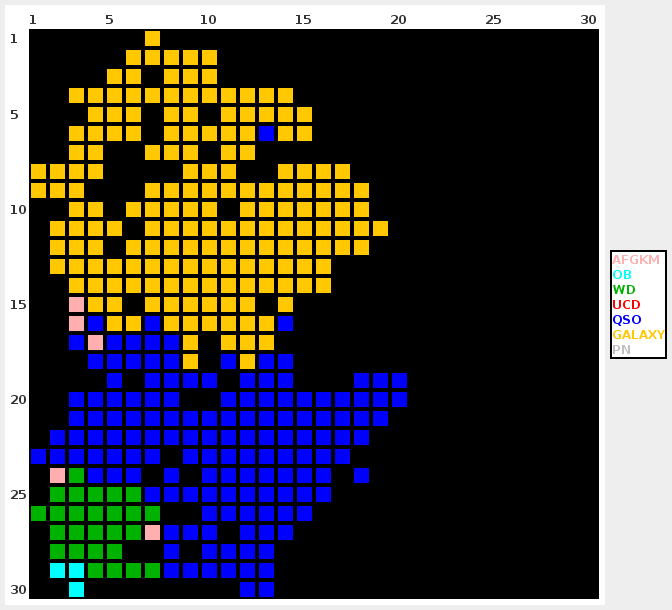} \label{fig:templateModes60} }
\caption{Identifications obtained for the SOM of SDSS outliers using Gaia simulations. The clusters receive a black color
when the distance between the outlier prototype and the corresponding template is above the established limit.}
  \label{fig:templates}
\end{figure}

As stated in previous sections, we cannot rely on supervised models to classify the unknown objects, since these would probably fail. However, we can still try to ``make a guess'' 
concerning the nature of the unknown objects. This can be achieved by means of distance-based models, such as a k-nearest neighbors (KNN) classifier. We have applied this method to label 
the units in the SOM built with the SDSS Outliers Library spectra, by retrieving the closer templates for each cluster prototype. The templates compiled with this purpose were obtained from the cluster prototypes in the SOM presented
in Section \ref{sect:oaalgorithm}. Fig. \ref{fig:templateModes} shows the results of the matching procedure, where each cluster in the SOM
of outliers was given a color depending on the class of the templates retrieved for it. From the figure, we can see that some regions 
in the map are filled with the same colors, thus receiving the same identification. For instance, the lower left corner of the map is dominated by white dwarfs (green clusters) and 
the lower center by quasars (blue units), whereas normal stars and galaxies
(pink and yellow clusters, respectively) are mostly located along the upper half of the map. Finally,
 ultra cool dwarfs are found in the upper right corner according to their very red colors.

\begin{figure}
\centering 

\subfigure[Prototype (blue) and best matching template (red) for unit at position (30,1)]
{\includegraphics[width=0.7\hsize] {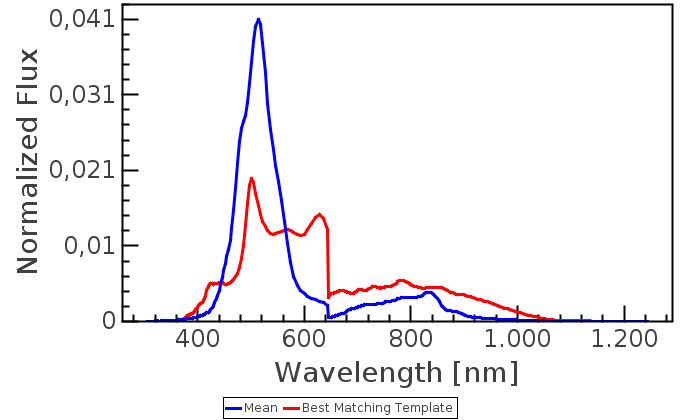}\label{fig:badfit} }

\subfigure[Prototype and best matching template for unit at position (28,4)]{   \includegraphics[width=0.7\hsize] {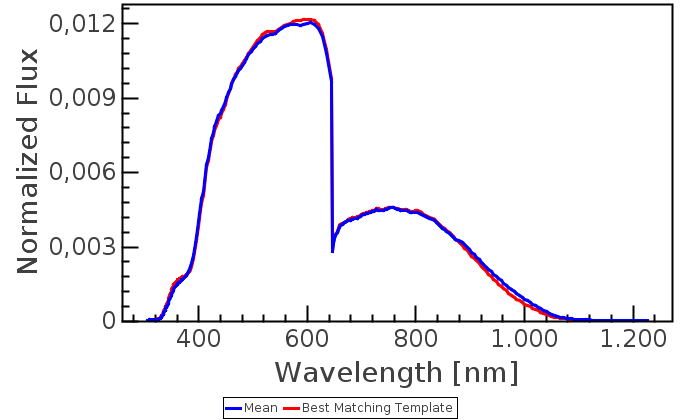} \label{fig:goodFit} }

\caption{Examples of fitness between the Gaia templates and two outlier SOM clusters.}
 \label{fig:labelfit}
\end{figure}

\begin{figure}
\centering
\subfigure[]{\includegraphics[width=0.45\hsize] {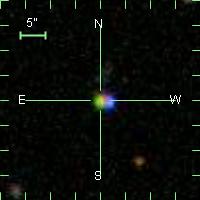} \label{fig:badImage1} }
\subfigure[]{\includegraphics[width=0.45\hsize] {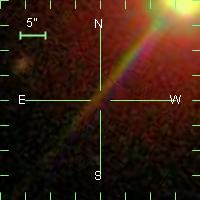} \label{fig:badImage2} }
\subfigure[]{\includegraphics[width=0.45\hsize] {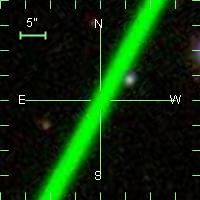} \label{fig:badImage3} }
\subfigure[]{\includegraphics[width=0.45\hsize] {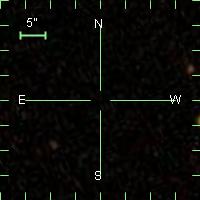} \label{fig:badImage4} }
\caption{Bad SDSS detections, populating the lower right region of the outliers SOM.}
  \label{fig:badImages}
\end{figure}

The previously described identification type can be useful as a first approach to study the nature of the SOM clusters, but one should be careful with the
 results, since these are rare/damaged objects. Therefore, the likelihood of the identifications should be studied. One way
 to obtain likelihoods is to look at the Euclidean distance obtained between the template and the outlier prototype that was obtained when the identification was performed.
  Fig. \ref{fig:labelfit} shows both the fitness obtained for neuron (30,1), which is the poorest one, and the fitness obtained for unit (5,4), where the prototype and the template are very close 
to each other. The distance can also be used to filter the regions in the map that are not likely to belong to
  known classes of astrophysical objects, as is shown in Figure \ref{fig:templates}. This filtering, together with the exploration of cluster
   prototypes and SDSS images, has allowed us to distinguish between common astrophysical objects and instrumental artifacts. For
   instance, the lower right region of the map, which was given a dark color in Figures \ref{fig:UMatrix} and \ref{fig:templateModes40},
    is populated by bad image detections, as the ones shown in Figure \ref{fig:badImages}.

\subsection{Cross-matching with external archives}\label{sect:crossmatching}
External archives can contribute to the identification process, since they contain additional information, including other wavelength ranges, 
imaging, other classifications, etc. Usually, cross-matching is performed by using a cone search in the sky, looking for objects within a certain radius. This functionality is 
already included in well-known tools such as Topcat or Aladin, which are integrated with the Virtual Observatory standards. In that sense, we can take advantage of the structure 
provided by the SOM to enhance the data exploration. 

\begin{figure*}
\centering
\subfigure[AGN]{\includegraphics[width=0.3\hsize] {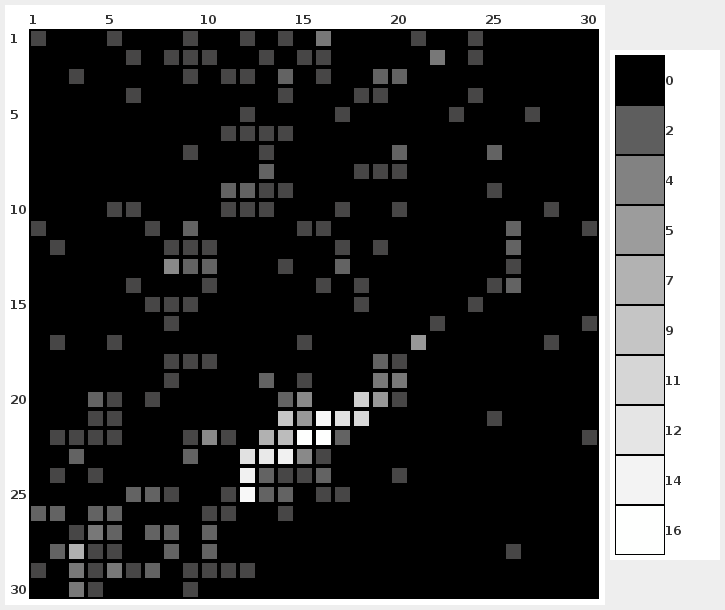} \label{fig:AGN-Hits} }
\subfigure[GALAXY]{\includegraphics[width=0.3\hsize] {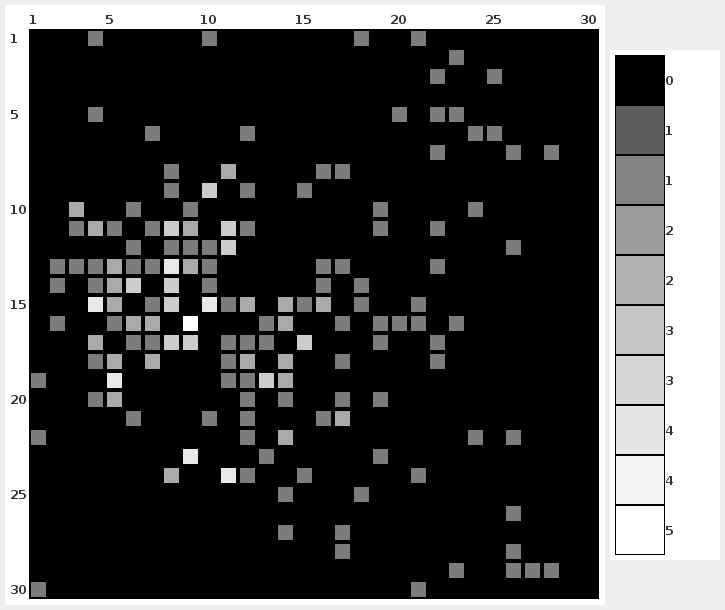} \label{fig:Galaxy-Hits} }
\subfigure[QSO]{\includegraphics[width=0.3\hsize] {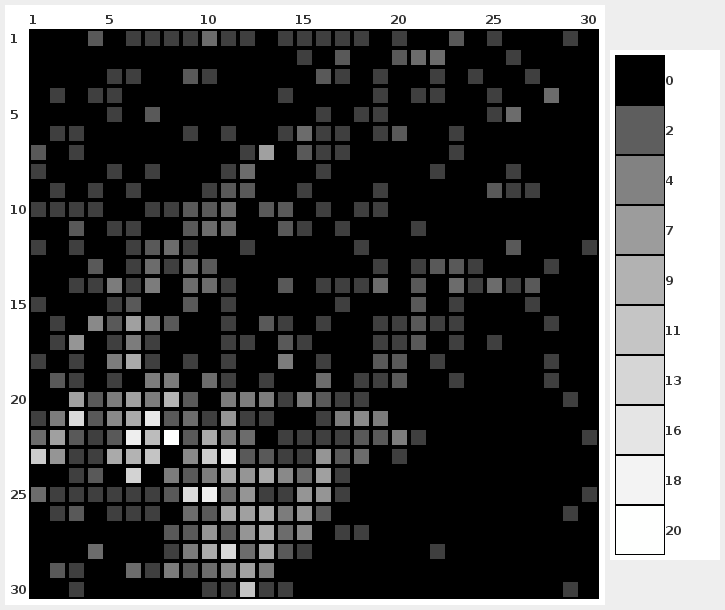} \label{fig:QSO-Hits} }
\subfigure[RADIO]{\includegraphics[width=0.3\hsize] {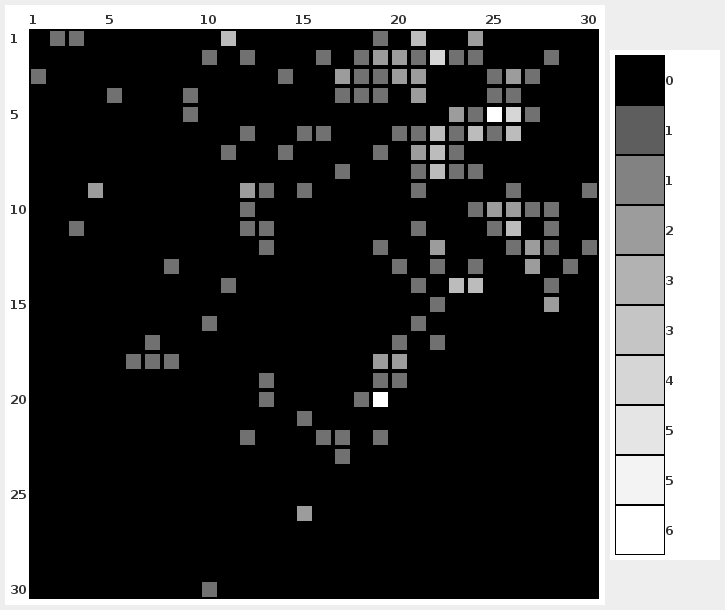} \label{fig:Radio-Hits} }
\subfigure[WD*]{\includegraphics[width=0.3\hsize] {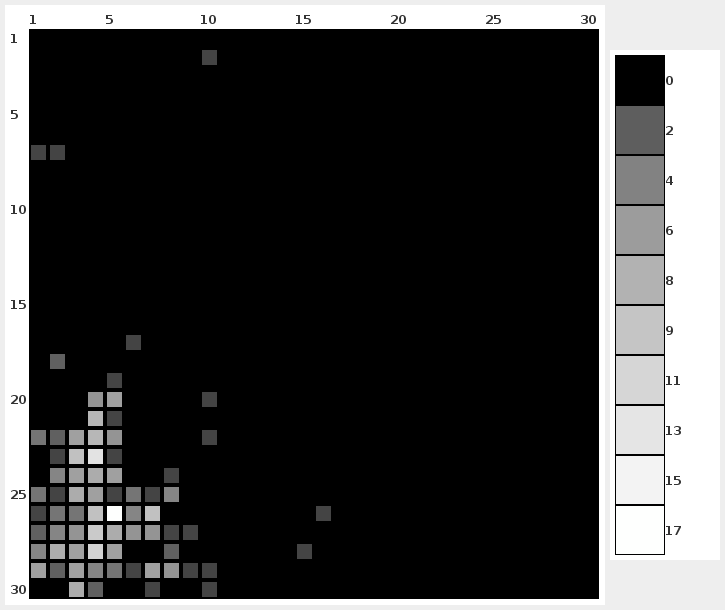} \label{fig:WD*-Hits} }
\subfigure[BROWND*]{\includegraphics[width=0.3\hsize] {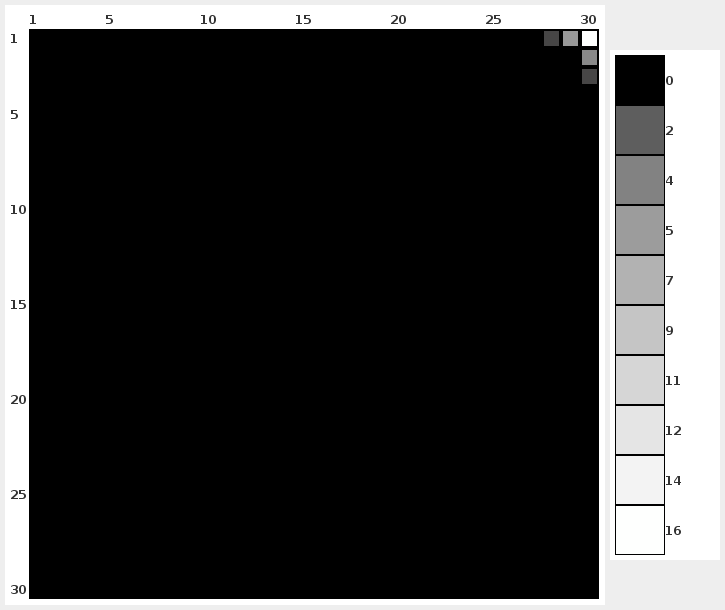} \label{fig:brownD*-Hits} }
\subfigure[LOW-MASS*]{\includegraphics[width=0.3\hsize] {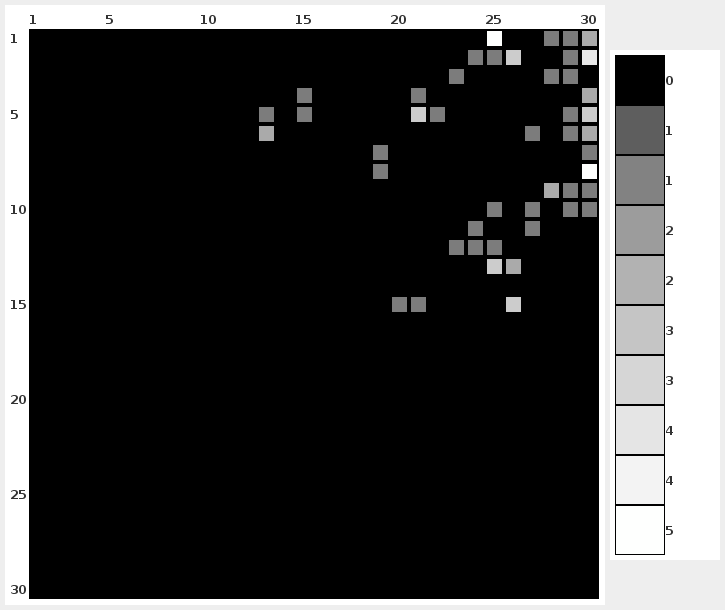} \label{fig:low-mass*-Hits} }
\subfigure[UNKNOWN]{\includegraphics[width=0.3\hsize] {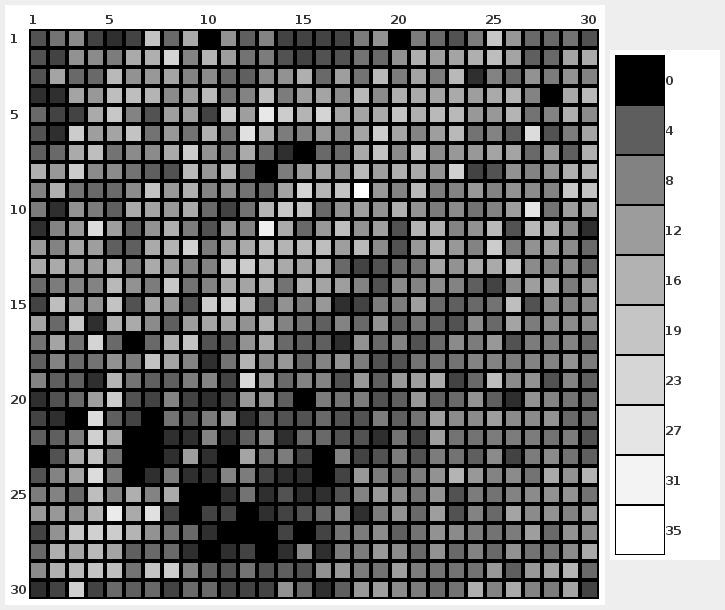} \label{fig:UNKNOWN-Hits} }

\caption{Diagram showing the number of hits in each SOM cluster obtained for the SDSS outliers library, for different SIMBAD classes (see Section 4.4 for details).}
  \label{fig:simbad-class-hits}
\end{figure*}

\begin{figure}
\centering
\subfigure[Simbad identifications for the outliers SOM]{\includegraphics[width=\hsize] {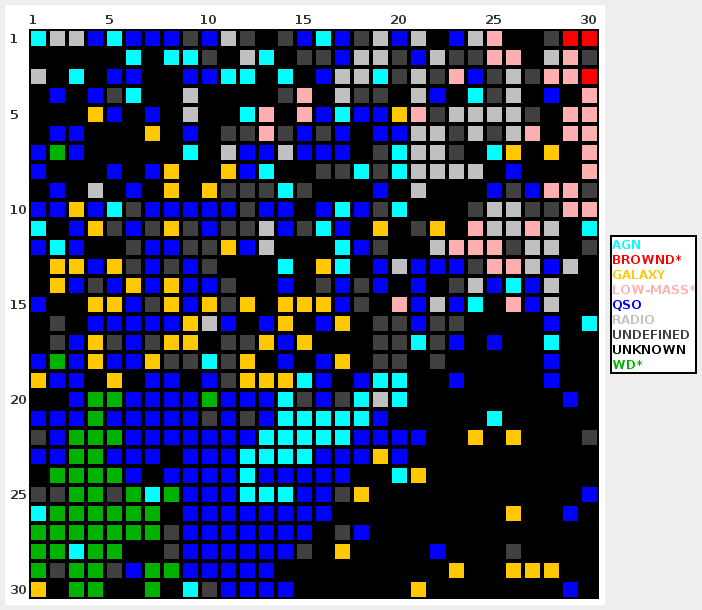} \label{fig:simbadModes} }
\subfigure[Class purity among identifications in the SIMBAD database for the outliers' SOM. A green color indicates that no object
in the cluster was identified ]{\includegraphics[width=\hsize] {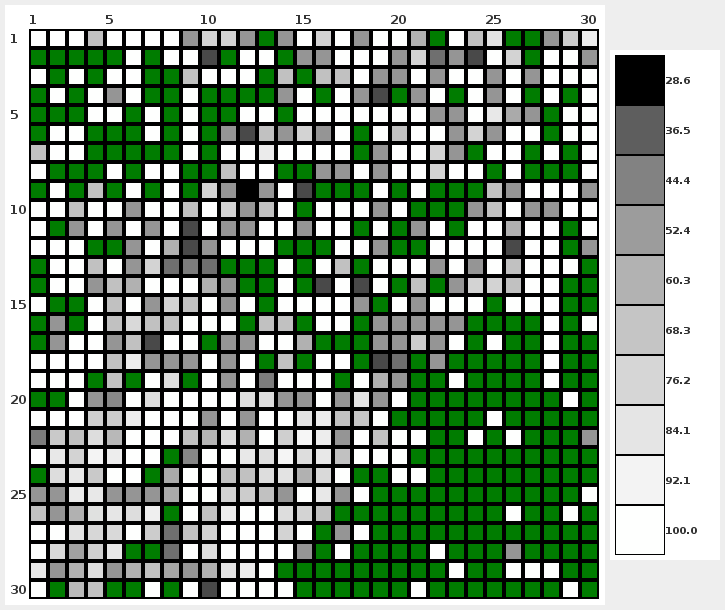} \label{fig:purity} }
\caption{Labeling of the map according to SIMBAD object types, see Sect \ref{sect:crossmatching} for details.}
 \label{fig:simbad-identification}
\end{figure}

We opted for the SIMBAD catalogue to perform cross-matching with the SDSS outliers, looking for further identifications. In this case, we retrieved those objects in SIMBAD 
within a radius of one arcsecond from every SDSS outlier, obtaining its SIMBAD type in case it exists. We obtained identifications among the following SIMBAD object types: 
AGN, Seyfert I galaxy, Seyfert II galaxy, BL-Lac object, galaxy, QSO, radio sources (in general), white dwarfs, brown dwarfs, and low-mass stars. For simplicity, SIMBAD classes 
AGN as well as Seyfert1, Seyfert2, and Bl-Lac objects were grouped together under the ``AGN'' label, which should be interpreted as active extragalactic objects excluding explicit 
SIMBAD identifications ``QSO''. It is to be observed that we expect almost no normal stars to form part of the SDSS outliers dataset. Figure \ref{fig:simbad-class-hits} shows the distribution of the 
retrieved SIMBAD identifications across the map. We can see that the different SIMBAD types are populating significantly separated regions in the SOM. It is remarkable that 
white dwarfs, quasars, and cool dwarfs are found in similar locations when identified by means of SIMBAD and Gaia simulations (see Fig. \ref{fig:templates}), 
which increases our confidence in the method. 
A  more compact view on the distribution of SIMBAD identifications is given in Figure \ref{fig:simbad-identification}. SOM clusters in Figure \ref{fig:simbadModes} receive a color in function of the most frequent SIMBAD identification. In the figure, black clusters do not have
 objects identified in SIMBAD and a grey color is assigned to clusters with a similar frequency among two or more classes. On the other hand, Figure
 \ref{fig:purity} displays, for each cluster, the purity (in percentage) of the most frequent SIMBAD class. Note that a green color is given
 to clusters without any SIMBAD identification. 

Apart from the distribution of SIMBAD identifications in the SOM, we performed a more formal evaluation through Confusion Matrices, which gives
a measure of how the different types of objects are being mixed in the SOM, as is shown in Table \ref{tab:confMatrixSIMBAD}. It can be observed that
the SOM is effective in classifying the SIMBAD types, especially considering the uncertainty introduced by the cross-matching and the
SIMBAD misclassification rate. In addition, Table \ref{tab:confMatrixSIMBAD} gives an idea of the discovery
 possibilities of our method. From 7898 objects without identifications in SIMBAD, 624 are candidates to be new WDs,
  1674 to be new QSOs and so on, following the percentages of the row corresponding to the UNKNOWN class.
 
\textbf{Table \ref{tab:confMatrixSIMBAD} can be compared with Table \ref{tab:confMatrixSIMBAD-noresp} and Table \ref{tab:confMatrixSIMBAD-HD}. Table
\ref{tab:confMatrixSIMBAD-noresp} shows the results obtained when the BP/RP data is divided by the response curve of the spectrophotometers,
while table \ref{tab:confMatrixSIMBAD-HD} shows the matrix when the SOM was computed from SDSS original high resolution spectra. In this way, we can evaluate
 the impact of the conversion from the SDSS spectra format to the Gaia BP/RP format. By analyzing these matrices, we can say that the algorithm behaves better
 working with BP/RP data. Even more, the BP/RP format without the application of any instrumental correction, chosen by CU8 for Apsis, 
 does not degrade the SOM performance. Provided that the outlier sources are very faint, the lack of resolution is compensated
 with a higher signal-to-noise ratio in the spectrophotometric data.}

\begin{table*}
\caption{Confusion matrix of the SOM, computed by means of he SDSS outliers semi-empirical library in Gaia spectrophotometers format. Identifications for the objects were obtained from SIMBAD.}             % title of Table
\label{tab:confMatrixSIMBAD}      % is used to refer this table in the text
\centering                          % used for centering table
{%
\newcommand{\mc}[3]{\multicolumn{#1}{#2}{#3}}
\begin{center}
\begin{tabular}{|r|r|r|r|r|r|r|r|r|r|}
\hline
 & \textbf{AGN} & \textbf{Galaxy} & \textbf{QSO} & \textbf{Radio} & \textbf{UNKNOWN} & \textbf{WD*} & \textbf{brownD*} & \textbf{low-mass*} & \textbf{UNDEFINED} \\ \hline
\textbf{AGN} & 60,5 & 1,14 & 10,5 & 2,97 & 0 & 9,82 & 0 & 0,46 & 15,07 \\ \hline
\textbf{Galaxy} & 4,52 & 47,96 & 18,55 & 2,26 & 0 & 1,36 & 0 & 0 & 25,34 \\ \hline
\textbf{QSO} & 4,55 & 1,01 & 79,05 & 1,32 & 0 & 3,95 & 0,10 & 0,20 & 9,82 \\ \hline
\textbf{Radio} & 4,37 & 1,09 & 7,65 & 58,47 & 0 & 0 & 0 & 0 & 28,42 \\ \hline
\textbf{UNKNOWN} & 6,75 & 8,88 & 17,61 & 7,42 & 32,53 & 7,90 & 0,22 & 6,05 & 12,65 \\ \hline
\textbf{WD*} & 2,69 & 0,34 & 4,71 & 0 & 0 & 85,86 & 0 & 0 & 6,40 \\ \hline
\textbf{brownD*} & 0 & 0 & 0 & 0 & 0 & 0 & 81,48 & 0 & 18,52 \\ \hline
\textbf{low-mass*} & 0 & 0 & 1,37 & 2,74 & 0 & 0 & 4,11 & 76,71 & 15,07 \\ \hline
\rowcolor{Gray}
 &  &  &  &  &  &  &  &  &  \\ \hline
\textbf{COUNT} & \multicolumn{1}{r|}{438} & \multicolumn{1}{r|}{221} & \multicolumn{1}{r|}{988} & \multicolumn{1}{r|}{183} & 7898 & \multicolumn{1}{r|}{297} & \multicolumn{1}{r|}{27} & \multicolumn{1}{r|}{73} & \multicolumn{1}{r|}{} \\ \hline
\end{tabular}
\end{center}
}%
\end{table*}

\begin{table*}
\caption{Confusion matrix of the SOM computed using roughly calibrated spectrophotometry. Identifications for the objects were obtained from SIMBAD. }             % title of Table
\label{tab:confMatrixSIMBAD-noresp}      % is used to refer this table in the text
\centering                          % used for centering table
{%
\newcommand{\mc}[3]{\multicolumn{#1}{#2}{#3}}
\begin{center}
\begin{tabular}{|r|r|r|r|r|r|r|r|r|r|}
\hline
 & \textbf{AGN} & \textbf{Galaxy} & \textbf{QSO} & \textbf{Radio} & \textbf{UNKNOWN} & \textbf{WD*} & \textbf{brownD*} & \textbf{low-mass*} & \textbf{UNDEFINED} \\ \hline
\textbf{AGN} & 52,05 & 0,91 & 21,23 & 1,60 & 0 & 9,36 & 0 & 0 & 14,84 \\ \hline
\textbf{Galaxy} & 4,98 & 53,85 & 18,55 & 0,90 & 0 & 0 & 0 & 0 & 21,72 \\ \hline
\textbf{QSO} & 5,87 & 1,42 & 77,13 & 1,32 & 0 & 3,95 & 0,20 & 0 & 10,12 \\ \hline
\textbf{Radio} & 9,29 & 1,09 & 13,11 & 50,82 & 0 & 0 & 0 & 0 & 25,68 \\ \hline
\textbf{UNKNOWN} & 6,19 & 7,55 & 20,23 & 6,96 & 35,41 & 6,82 & 0,34 & 3,30 & 13,18 \\ \hline
\textbf{WD*} & 2,02 & 0,34 & 18,86 & 0 & 0 & 73,74 & 0 & 0 & 5,05 \\ \hline
\textbf{brownD*} & 0 & 0 & 0 & 0 & 0 & 0 & 81,48 & 0 & 18,52 \\ \hline
\textbf{low-mass*} & 0 & 0 & 5,48 & 6,85 & 0 & 0 & 4,11 & 54,79 & 28,77 \\ \hline
\rowcolor{Gray}
 &  &  &  &  &  &  &  &  &  \\ \hline
\textbf{COUNT} & \multicolumn{1}{r|}{438} & \multicolumn{1}{r|}{221} & \multicolumn{1}{r|}{988} & \multicolumn{1}{r|}{183} & 7898 & \multicolumn{1}{r|}{297} & \multicolumn{1}{r|}{27} & 73 &  \\ \hline
\end{tabular}
\end{center}
}%
\end{table*}

\begin{table*}
\caption{Confusion matrix of the SOM computed using SDSS full resolution spectra, classified as UNKNOWN.
 Identifications for the objects were obtained from SIMBAD. }             % title of Table
\label{tab:confMatrixSIMBAD-HD}      % is used to refer this table in the text
\centering                          % used for centering table
{%
\newcommand{\mc}[3]{\multicolumn{#1}{#2}{#3}}
\begin{center}
\begin{tabular}{|r|r|r|r|r|r|r|r|r|r|}
\hline
 & \textbf{AGN} & \textbf{Galaxy} & \textbf{QSO} & \textbf{Radio} & \textbf{UNKNOWN} & \textbf{WD*} & \textbf{brownD*} & \textbf{low-mass*} & \textbf{UNDEFINED} \\ \hline
\textbf{AGN} & 52,51 & 0,68 & \multicolumn{1}{r|}{21} & 1,83 & 0 & 7,76 & 0 & 0 & 16,21 \\ \hline
\textbf{Galaxy} & 3,62 & 42,08 & 30,77 & 1,36 & 0 & 0 & 0 & 0 & 22,17 \\ \hline
\textbf{QSO} & 5,67 & 0,81 & 77,94 & 1,21 & 0 & 2,94 & 0,10 & 0,30 & 11,03 \\ \hline
\textbf{Radio} & 7,65 & 1,09 & 9,29 & 53,01 & 0 & 0 & 0 & 2,19 & 26,78 \\ \hline
\textbf{UNKNOWN} & 5,58 & 7,61 & 21,25 & 7,50 & 34,89 & 6,31 & 0,38 & 3,32 & 13,17 \\ \hline
\textbf{WD*} & 1,35 & 0 & 20,54 & 0 & 0 & 68,35 & 0 & 0 & 9,76 \\ \hline
\textbf{brownD*} & 0 & 0 & 0 & 0 & 0 & 0 & 85,19 & 3,70 & 11,11 \\ \hline
\textbf{low-mass*} & 0 & 0 & 5,48 & 9,59 & 0 & 0 & 5,48 & 53,42 & 26,03 \\ \hline
\rowcolor{Gray}
 &  &  &  &  &  &  &  &  &  \\ \hline
\textbf{COUNT} & \multicolumn{1}{r|}{438} & \multicolumn{1}{r|}{221} & \multicolumn{1}{r|}{988} & \multicolumn{1}{r|}{183} & 7898 & \multicolumn{1}{r|}{297} & \multicolumn{1}{r|}{27} & \multicolumn{1}{r|}{73} & \multicolumn{1}{r|}{} \\ \hline
\end{tabular}
\end{center}
}%
\end{table*}

\section{Discussion and future developments}\label{sect:conclusions}
As a precise, large, and complete survey, Gaia is raising enormous expectation from the astrophysical community. In order to process such a tremendous amount of data, automated
 specialized analysis tools are being developed by the Gaia DPAC, with the aim of classifying objects and estimating their astrophysical parameters. A set of standardized classification
 labels are used in order to enable the supervised classification of approximately 95\% of the observed sources. In this sense, an enormous 
amount of outliers is expected (of the magnitude of $5*10^{7}$), which will be detected as objects that cannot be reliably related to any of the predefined categories, either because
 their observations are complex (several superposed objects, instrumental/calibration errors, signals with high noise levels, etc.) or because they belong to a new class of objects that 
is rarely found. The purpose of the OutlierAnalysis Gaia DPAC group is to prepare algorithms for analyzing such objects by means of unsupervised classification techniques.

Our work presents and discusses the application of the Outlier Analysis algorithms to a significant set of spectra from the well-known SDSS survey, with the aim of characterizing the 
spectroscopic classification outliers. Specifically designed SOMs were used to compress the dataset in an euclidean distance optimal representation. This simplifies 
the posterior analysis, since we can perform complex operations on the obtained clusters. Furthermore, the SOM algorithms project the dataset onto a two- dimensional grid where 
the topological relations are preserved, allowing us to easily visualize the dataset distribution, in order to find clusters and outliers, and facilitating data exploration and knowledge 
discovery.
 
With the identifications obtained by means of Gaia spectrophotometric simulations and by the retrieval of external data via the SDSS SkyServer and the SIMBAD database, we were able 
to identify the nature of clusters populating some of the SOM regions, including classes of unknown objects whose spectra were not included among the templates in the current Gaia 
simulation (such as some types of AGN as BL-lac objects). We also identified, among the outlying regions, several sources of errors, such as poor photometric detections or 
incomplete spectra, and some objects of an uncommon nature that require further identification.

The results shown in this work demonstrate that the proposed method can effectively assist researchers in making a distinction among source candidates
 to complete the training sets of supervised classification algorithms, faint/damaged objects, and new classes of astronomical objects. Our expectation
 is that this will help the process of data processing of the upcoming Gaia dataset, finding, as soon as possible, systematic errors in the instrumental or in the pipeline procedures. Additionally, 
we hope that it will be possible to detect new objects unseen before, with the help of the methods presented here and of the astronomical community. However, the Gaia mission will 
bring new challenges, such as the processing of large amounts of data, high
 extinction levels, etc., that will be addressed in the near future.

There is still work to be done in order to extend the Outlier Analysis functionality. The present paper has shown a process of identification that makes use of spectrophotometric templates and semiautomated cross-matching of sources. This identification procedure will be extended in the future by incorporating additional 
data, such as Gaia astrometry and photometry, as well as measured spectral features, integrating an Expert System (see \cite{RandallKBS}) that would make inference with all available 
internal Gaia data. On the other hand, the process of cross-matching could take into account other characteristics apart from positions in the sky, such as for instance photometric similarities. 
Also, the cross-matching method will be further automated, by building robots that could navigate the Web looking for good identifications
 in surveys such as LSST, LAMOST, or VISTA. Finally, interesting objects could be proposed for follow-up programmes and identification by new ad-hoc observations.

\begin{acknowledgements}
This work was supported by the Spanish \emph{MINECO} – \emph{FEDER}
 through Grants AYA2009-14648-C02-02, AYA2009-14648-C02-01, and
 \emph{CONSOLIDER} CSD2007-00050. The GOG simulations were run on the supercomputer
  MareNostrum at the Barcelona Supercomputing Center – Centro Nacional de Supercomputación. 
  In addition, we would like to acknowledge the support of the Italian Space Agency, through ASI contract
   I/058/10/0, and that of the German space agency (DLR).
\end{acknowledgements}

\bibliographystyle{aa} % style aa.bst
\bibliography{SDSSOutlierAnalysis.Gaia} % your references Yourfile.bib

\end{document}